\documentclass{appolb}%
\usepackage{epsfig}
\usepackage{amsmath}
\usepackage{graphicx}%
\usepackage{amsfonts}%
\usepackage{amssymb}
\def\runhead{MFV in supersymmetric theories -- Christopher Smith}
\def\appheadit{\small\scshape}

\begin{document}

\title{Minimal flavor violation in supersymmetric theories\thanks{Presented at the Flavianet Topical Workshop \textit{Low energy constraints on extensions of the Standard Model, July 23-27, Kazimierz, Poland}.
\\ Many thanks to the organizers for their kind invitation to this pleasant workshop. Work supported by EU contract MRTN-CT-2006-035482 and project C6 of the DFG Research Unit SFT-TR9 \textit{Computergestützte Theoretische Teilchenphysik}.}}

\author{Christopher Smith \address{Institut f\"{u}r Theoretische Teilchenphysik, Karlsruhe Institute of Technology, D-76128 Karlsruhe, Germany}}

\maketitle

\begin{abstract}
The Minimal Flavor Violation (MFV) hypothesis is presented in the context of the MSSM. Its fundamental principles are introduced and motivated, and its phenomenological consequences for FCNC and CP-violating observables as well as for R-parity violating processes are briefly described.

\end{abstract}


\section{Introduction}

The aim of the MFV hypothesis is to reconcile two contradictory facts. On one hand, tremendous advances from flavor factories have confirmed the Standard Model (SM) flavor structures to an impressive precision, in both the quark sector through $K$ and $B$ physics observables, and in the lepton sector where tight bounds on flavor transitions have been obtained. On the other hand, most New Physics (NP) models extend the SM flavor sector. With new degrees of freedom at or below the TeV scale, so as to avoid hierarchy problems, NP signals should have been seen in flavor physics. 

Thus, the NP flavor structures cannot be generic but are necessarily fine-tuned to pass all experimental constraints. Our purpose here is to describe how MFV manages to naturally achieve these adjustments in the flavor sectors of the MSSM, the minimal supersymmetric extension to the SM. Before that, in the next section, we analyze in some details the flavor structures of both the SM and MSSM using the language of the flavor symmetry and its spurions. This language is central to the formulation of the two conditions for MFV, described in Sec.~3. Finally, the phenomenological implications are briefly reviewed in Sec.~4.

\section{The flavor symmetry and its uses}

In the SM and MSSM, the three generations of matter fields can be freely and independently redefined for each matter species without affecting the gauge sector, which thus has the symmetry~\cite{ChivukulaG87}%
\begin{equation}
G_{F}\equiv U(3)^{5}=U\left( 3\right)_{Q}\times U\left( 3\right)_{U}\times U\left( 3\right)_{D}\times U\left( 3\right)_{L}\times U\left( 3\right)_{E}\;,
\end{equation}
where $Q=(u_{L},d_{L})$, $U=u_{R}^{\dagger}$, $D=d_{R}^{\dagger}$, $L=(\nu_{L},e_{L})$, and $E=e_{R}^{\dagger}$ stand for fields or superfields. This symmetry is broken differently in the SM and MSSM, so let us analyze each one in turn.

\subsubsection*{SM flavor structures}

In the SM, $G_{F}$ is broken only by the Yukawa couplings, which generate fermion masses and mixing after electroweak symmetry breaking (EWSB):%
\begin{equation}
\mathcal{L}_{\mathrm{SM}}\ni U\mathbf{Y}_{u}QH+D\mathbf{Y}_{d}QH^{\ast}+E\mathbf{Y}_{e}LH^{\ast}\;. \label{Yuk}%
\end{equation}
The $\mathbf{Y}_{u,d,e}$ have peculiar structures: the masses are highly hierarchical, while the only mixing ($\mathbf{Y}_{e}$ can always be diagonalized) is described by the hierarchical CKM matrix. To analyze the phenomenological consequences of this, let us use the flavor symmetry and adopt the spurion language.

We start by observing that the whole SM Lagrangian becomes invariant under $G_{F}$ if we promote the Yukawa couplings to spurions transforming as%
\begin{equation}
\mathbf{Y}_{u}\rightarrow g_{U}\mathbf{Y}_{u}g_{Q}^{\dagger},\;\;\mathbf{Y}_{d}\rightarrow g_{D}\mathbf{Y}_{d}g_{Q}^{\dagger},\;\;\mathbf{Y}_{e}\rightarrow g_{E}\mathbf{Y}_{e}g_{L}^{\dagger}\;,
\end{equation}
where $g_{X}\in U(3)_{X}$. This makes Eq.~(\ref{Yuk}) invariant under $Q\rightarrow g_{Q}Q$, $U\rightarrow Ug_{U}^{\dagger}$,... , and the $U(3)^{5}$ symmetry becomes exact. Note that by spurions one understands non-dynamical (constant) fields with definite transformation properties, i.e. completely artificial theoretical constructs. Even so, we now have a powerful tool at hand. Once the SM Lagrangian is $G_{F}$-invariant, even if artificially, so is any SM amplitude. Hence, using only the symmetry, we can guess the flavor structure of any amplitude. These predictions even become quantitative by freezing $\mathbf{Y}_{u,d,e}$ back to their physical values: $v\mathbf{Y}_{u}=\mathbf{m}_{u}V$,$\ v\mathbf{Y}_{d}=\mathbf{m}_{d}$, and $v\mathbf{Y}_{e}=\mathbf{m}_{e}$, with $\mathbf{m}_{u,d,e}$ the diagonal mass matrices and $v$ the Higgs vacuum expectation value (VEV). The CKM matrix $V$ is put in $\mathbf{Y}_{u}$ such that the down quarks are immediately mass eigenstates.

Let us give a few examples to show that the spurion language is just an innocuous theoretical trick, and no approximation is made on the SM flavor sector. First, consider the $Z$ penguin transition $d_{L}^{I}\rightarrow d_{L}^{J}Z$, described by the operator $\bar{Q}^{I}\gamma^{\mu}Q^{J}\,\,H^{\dagger}D_{\mu}H$ (with $H^{\dagger}D_{\mu}H\rightarrow v^{2}Z_{\mu}$ after EWSB). For it to be flavor-violating and $G_{F}$-symmetric, its simplest form must be%
\begin{equation}
\mathcal{O}_{Z}^{\mathrm{SM}}\sim M_{W}^{-2}\,(\bar{Q}\mathbf{Y}_{u}^{\dagger}\mathbf{Y}_{u}\gamma^{\mu}Q)\,\,H^{\dagger}D_{\mu}H\;. \label{SMz}%
\end{equation}
Freezing the spurions, the transitions $d_{L}^{I}\rightarrow d_{L}^{J}Z$ are tuned by $v^{2}(\mathbf{Y}_{u}^{\dagger}\mathbf{Y}_{u})^{IJ}\sim m_{t}^{2}V_{3I}^{\ast}V_{3J}$, which is precisely what would give an exact computation of the FCNC loop. The same exercise can be done for $d_{R}^{I}\rightarrow d_{R}^{J}Z$. The simplest flavor-violating operator automatically generates the chirality flips: it has the structure $D\mathbf{Y}_{d}\mathbf{Y}_{u}^{\dagger}\mathbf{Y}_{u}\mathbf{Y}_{d}^{\dagger}\gamma^{\mu}\bar{D}$ and is thus suppressed by $m_{d^{I}}m_{d^{J}}/v^{2}$ compared to $\mathcal{O}_{Z}^{\mathrm{SM}}$. Finally, in the lepton sector, the only spurion, $\mathbf{Y}_{e}$, is diagonal, hence there can be no flavor transitions like $\ell^{I}\rightarrow\ell^{J}\gamma$.

\subsubsection*{MSSM flavor structures}

In the MSSM, in addition to the Yukawa couplings, all the sfermion soft-breaking terms, remnants of the unknown SUSY breaking mechanism, are new flavor couplings. Therefore, to create the $U(3)^{5}$ flavor symmetry, many more couplings have to be promoted to spurions. But doing this opens up new options to induce flavor transitions, for example as%
\begin{align}
\mathcal{L}_{\mathrm{Soft}}  &  \ni\tilde{Q}^{\dagger}\mathbf{m}_{Q}^{2}\tilde{Q}\;\;\rightarrow\;\;\mathcal{O}_{Z}^{\mathrm{MSSM}}\sim M_{SUSY}^{-4}\,(\bar{Q}\mathbf{m}_{Q}^{2}\gamma^{\mu}Q)\,\,H_{u}^{\dagger}D_{\mu}H_u\;,\label{SUSYz}\\
\mathcal{L}_{\mathrm{Soft}}  &  \ni\tilde{L}^{\dagger}\mathbf{m}_{L}^{2}\tilde{L}\;\;\rightarrow\;\;\mathcal{O}_{\gamma}^{\mathrm{MSSM}}\sim M_{SUSY}^{-4}\,(E\mathbf{Y}_{e}\mathbf{m}_{L}^{2}\sigma^{\mu\nu}L)\,H_{d}\,F_{\mu\nu}\;, \label{SUSYg}%
\end{align}
where $H_{u,d}$ are the MSSM Higgses with VEV $v_{u,d}$. If the SUSY scale $M_{SUSY}$ is close to the SM scale and if the new spurions are completely generic, e.g. $\mathbf{m}_{Q,L}^{2}/M_{SUSY}^{2}\sim \mathcal{O}(1)$, flavor transitions can occur at unacceptable rates. To avoid this, all the extra spurions of the MSSM must be close to those of the SM, e.g. $\mathbf{m}_{Q}^{2}\sim\mathbf{Y}_{u}^{\dagger}\mathbf{Y}_{u}$ and $\mathbf{m}_{L}^{2}\sim\mathbf{Y}_{e}^{\dagger}\mathbf{Y}_{e}$. This apparent incompatibility between naturality, low-scale SUSY, and FCNC constraints is called the MSSM flavor puzzle.

This is not the full story yet, because the conservation of baryon ($\mathcal{B}$) and lepton ($\mathcal{L}$) numbers is not automatic in the MSSM (in the SM, $\mathbf{Y}_{u,d,e}$ are the only renormalizable flavor couplings). Indeed, the superpotential can contain the terms (the corresponding soft-breaking terms are understood)%
\begin{equation}
\mathcal{W}_{RPV}\ni\boldsymbol{\mu}^{\prime}LH_{d}+\boldsymbol{\lambda}LLE+\boldsymbol{\lambda}^{\prime}LQD+\boldsymbol{\lambda}^{\prime\prime}UDD\;,\label{RPV}%
\end{equation}
which induce new transitions like proton decay. The experimental constraint $\tau_{p^{+}}\gtrsim10^{31}$ years~\cite{PDG} implies that these couplings must be so small that one usually introduces by hand a new symmetry, R-parity, to forbid them altogether. Still, the R-parity violating (RPV) $\Delta\mathcal{B}$ and $\Delta\mathcal{L}$ interactions are flavored, and the mechanism solving the other flavor puzzles, i.e. those due to the quark and lepton flavor transitions, could well solve this one also.

\subsubsection*{Seesaw mechanism}

For a complete flavor sector, neutrino masses have to be introduced. To avoid unnaturally small parameters, we do this with a seesaw mechanism of type I~\cite{Seesaw}. Three right-handed neutrino fields (superfields) $N$, with the couplings $\tfrac{1}{2}N^{T}\mathbf{M}_{R}N+N\mathbf{Y}_{\nu}LH_{(u)}$, are added to the SM Lagrangian (MSSM superpotential). The phenomenological consequences can again be analyzed with the help of the spurion language. Though the flavor symmetry is extended to $G_{F}\times U(3)_{N}$, the heavy $N$ fields never occur at low-energy and only the spurion combinations which are singlets under $U(3)_{N}$ are relevant. Expanding in the inverse mass-matrix $\mathbf{M}_{R}$, these are~\cite{CiriglianoGIW05}%
\begin{equation}
\mathbf{Y}_{\nu}^{\dagger}\mathbf{Y}_{\nu},\;\mathbf{Y}_{\nu}^{T}(\mathbf{M}_{R}^{-1})\mathbf{Y}_{\nu},\;\;\;\mathbf{Y}_{\nu}^{T}(\mathbf{M}_{R}^{-1})(\mathbf{M}_{R}^{-1})^{\ast}\mathbf{Y}_{\nu}^{\ast},\;...\label{Eq5}%
\end{equation}
In the charged lepton mass-eigenstate basis, the $\nu_{L}$ Majorana mass term is%
\begin{equation}
v_{(u)}\boldsymbol{\Upsilon}  _{\nu}\equiv v_{(u)}^{2}\mathbf{Y}_{\nu}^{T}(\mathbf{M}_{R}^{-1})\mathbf{Y}_{\nu}=U^{\ast}\mathbf{m}_{\nu}U^{\dagger}\;,\label{numass}%
\end{equation}
with $U$ the PMNS matrix. If $\mathbf{M}_{R}\sim\mathcal{O(}10^{13})$ GeV, $\mathbf{Y}_{\nu}$ can be of $\mathcal{O}(1)$ given the measured $\nu_{L}$ masses. For such a large $\mathbf{M}_{R}$, the only other relevant spurion combination is $\mathbf{Y}_{\nu}^{\dagger}\mathbf{Y}_{\nu}$. Note that knowing $\mathbf{m}_{\nu}$ and $U$, and even assuming $\mathbf{M}_{R}=M_{R}\mathbf{1}$, some freedom remains for $\mathbf{Y}_{\nu}^{\dagger}\mathbf{Y}_{\nu}$, see e.g.~\cite{CasasI01}.

Being non-diagonal, $\mathbf{Y}_{\nu}^{\dagger}\mathbf{Y}_{\nu}$ allows for  lepton flavor violating (LFV) transitions. For example, the operator $E\mathbf{Y}_{e}\mathbf{Y}_{\nu}^{\dagger}\mathbf{Y}_{\nu}\sigma^{\mu\nu}LF_{\mu\nu}$ can induce $\ell^{I}\rightarrow\ell^{J}\gamma$. However, the situation is different in the SM and MSSM. In the SM, any sensitivity to $\mathbf{Y}_{\nu}$ occurs through exchanges of $N$. Thus, $\ell^{I}\rightarrow\ell^{J}\gamma$ is tuned by $\mathbf{Y}_{\nu}^{\dagger}\mathbf{Y}_{\nu}/M_{R}\sim\mathcal{O}(\mathbf{m}_{\nu})$ and is too small to be observed. In the MSSM, the presence of the $N$ superfields affects the slepton mass terms at the high-scale, which then propagate the effects down to the low scale~\cite{BorzumatiM86}. With $\mathbf{m}_{L}^{2}/M_{SUSY}^{2}\sim\mathbf{Y}_{\nu}^{\dagger}\mathbf{Y}_{\nu}\sim\mathcal{O}(1)$ in Eq.~(\ref{SUSYg}), LFV transitions may end up too large compared to experimental bounds, generating a new flavor puzzle.

\section{The minimal flavor violation hypothesis}

The MFV hypothesis is defined by two conditions. The first specifies how the flavor couplings are to be constructed, and the second requires all the free parameters to be natural. Let us describe each of them in detail.

\subsubsection*{Construction principle}

The first condition for MFV is simple to express in the spurion language:\textit{ }all the flavor couplings are required to be invariant under $G_{F}$, but only the spurions needed to account for the fermion masses and mixings are allowed. This is clearly a minimal breaking of $G_{F}$, since anything less would be insufficient to reproduce the well-known fermionic flavor structures.

\begin{figure}[t]
\centering   \includegraphics[width=0.98\textwidth]{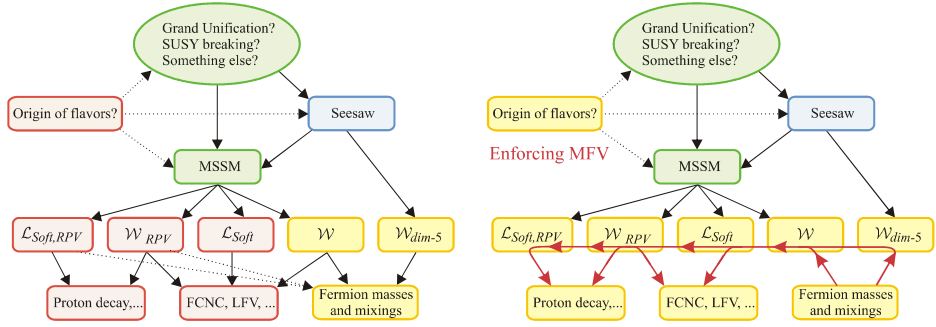}
\caption{As a first step towards solving the flavor puzzles, only the elementary flavor structures required to generate the hierarchical fermion masses and mixings are allowed. The symmetry principle then dictates how all the other MSSM flavor couplings can be constructed in terms of these minimal breaking terms.}
\label{Fig1}%
\end{figure}

MFV is a statement about the mechanism at the origin of the flavor structures, see Fig.~\ref{Fig1}. Indeed, no matter the physics at the high-scale, it is always possible to promote its fundamental flavor structures to spurions to make it (artificially) flavor symmetric, exactly as described in Sec.~2 for the SM and MSSM. What MFV asks is for these spurions to be necessary and sufficient to generate the Yukawa couplings $\mathbf{Y}_{u,d,e}$ (here and in the following, the neutrino spurions $\mathbf{Y}_{\nu}$ and $\mathbf{M}_{R}$ are understood). But if the high-scale spurions are equivalent to the Yukawa couplings, we do not need their precise forms and can simply trade them for $\mathbf{Y}_{u,d,e}$. All the low-scale MSSM flavor couplings can then be reconstructed entirely out of $\mathbf{Y}_{u,d,e}$. For example, the squark mass term $\mathbf{m}_{Q}^{2}$ necessarily takes the form~\cite{HallR90,DambrosioGIS02}
\begin{equation}
\mathbf{m}_{Q}^{2}=m_{0}^{2}(z_{1}\mathbf{1}+z_{2}\mathbf{Y}_{u}^{\dagger}\mathbf{Y}_{u}+z_{3}\mathbf{Y}_{d}^{\dagger}\mathbf{Y}_{d}+z_{4}\{\mathbf{Y}_{d}^{\dagger}\mathbf{Y}_{d},\mathbf{Y}_{u}^{\dagger}\mathbf{Y}_{u}\}+\,...)\;, \label{MFVexp}%
\end{equation}
with the SUSY-breaking scale $m_{0}\sim1\,$TeV. Let us stress that even if MFV uses the Yukawa couplings as building blocks, they are not given any new physical content; they are just the remnants of the flavor structures of some high-energy theory beyond the MSSM. MFV does not specify anything about the high-energy dynamics, in particular it does not require $G_{F}$ to be realized exactly (instead of artificially) at any scale.

The MFV series is always finite, because of the Cayley-Hamilton matrix identity for a generic $3\times3$ matrix $\mathbf{X}$
\begin{equation}
\mathbf{X}^{3}-\langle\mathbf{X\rangle X}^{2}+\tfrac{1}{2}\mathbf{X}(\langle\mathbf{X\rangle}^{2}-\langle\mathbf{X}^{2}\mathbf{\rangle)}-\det\mathbf{X}=0\;,\label{CH1}%
\end{equation}
where $\langle\mathbf{X\rangle}$ denotes the trace of $\mathbf{X}$. This identity holds at the spurion level. Enforcing it leaves $17$ hermitian terms in the expansion of each soft-breaking term~\cite{NikolidakisS07,ColangeloNS08,MercolliS09}. Once the spurions are frozen to their physical values, the number of terms is further reduced using the approximate identities $(\mathbf{Y}_{u}^{\dagger}\mathbf{Y}_{u})^{2}%
\sim\mathbf{Y}_{u}^{\dagger}\mathbf{Y}_{u}$ and similarly for $\mathbf{Y}_{d}$ and $\mathbf{Y}_{e}$. This leaves only five (nine) hermitian terms per squark (slepton) soft-breaking term, for example:%
\begin{align}
\mathbf{m}_{Q}^{2} &  =m_{0}^{2}(a_{1}\mathbf{1}+a_{2}\mathbf{Y}_{u}^{\dagger
}\mathbf{Y}_{u}+a_{3}\mathbf{Y}_{d}^{\dagger}\mathbf{Y}_{d}\nonumber\\
&  \;\;\;\;\;+a_{4}\{\mathbf{Y}_{u}^{\dagger}\mathbf{Y}_{u},\mathbf{Y}_{d}^{\dagger}\mathbf{Y}_{d}\}+a_{5}i[\mathbf{Y}_{u}^{\dagger}\mathbf{Y}_{u},\mathbf{Y}_{d}^{\dagger}\mathbf{Y}_{d}])\,,\nonumber\\
\mathbf{m}_{L}^{2} &  =m_{0}^{2}(b_{1}\mathbf{1}+b_{2}\mathbf{Y}_{e}^{\dagger}\mathbf{Y}_{e}+b_{3}\mathbf{Y}_{\nu}^{\dagger}\mathbf{Y}_{\nu}+b_{4}(\mathbf{Y}_{\nu}^{\dagger}\mathbf{Y}_{\nu})^{2}+b_{5}\{\mathbf{Y}_{e}^{\dagger}\mathbf{Y}_{e},\mathbf{Y}_{\nu}^{\dagger}\mathbf{Y}_{\nu}\}\nonumber\\
&  \;\;\;\;\;+b_{6}\mathbf{Y}_{\nu}^{\dagger}\mathbf{Y}_{\nu}\mathbf{\mathbf{Y}}_{e}^{\dagger}\mathbf{\mathbf{Y}}_{e}\mathbf{Y}_{\nu}^{\dagger}\mathbf{Y}_{\nu}+b_{7}i[\mathbf{Y}_{e}^{\dagger}\mathbf{Y}_{e},\mathbf{Y}_{\nu}^{\dagger}\mathbf{Y}_{\nu}]+b_{8}i[\mathbf{Y}_{e}^{\dagger}\mathbf{Y}_{e},(\mathbf{Y}_{\nu}^{\dagger}\mathbf{Y}_{\nu})^{2}]\nonumber\\
&  \;\;\;\;\;+b_{9}i((\mathbf{Y}_{\nu}^{\dagger}\mathbf{Y}_{\nu})^{2}\mathbf{Y}_{e}^{\dagger}\mathbf{Y}_{e}\mathbf{Y}_{\nu}^{\dagger}\mathbf{Y}_{\nu}-\mathbf{Y}_{\nu}^{\dagger}\mathbf{Y}_{\nu}\mathbf{\mathbf{Y}}_{e}^{\dagger}\mathbf{\mathbf{Y}}_{e}(\mathbf{Y}_{\nu}^{\dagger}\mathbf{Y}_{\nu})^{2})\;.\label{HermQ}%
\end{align}
(The expansions for RPV couplings are slightly different, see Sec.~4.) How these structures can solve the flavor puzzles depend on the coefficients $a_{i}$ and $b_{i}$, to which we now turn our attention.

\subsubsection*{Naturality principle}

\begin{figure}[t]
\centering   \includegraphics[width=0.98\textwidth]{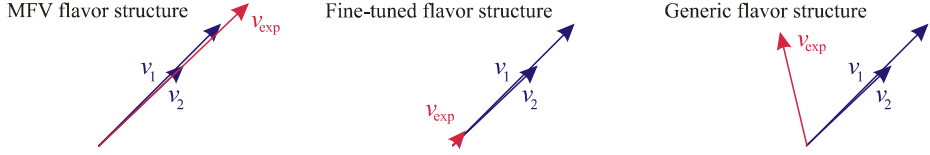}
\caption{The three possible MSSM flavor structures. The vectors $v_{1,2}$ represent the basis made of the $G_{F}$ symmetric terms of~(\ref{HermQ}). Expressed in this basis, $v_{\exp}$, obtained from flavor experiments, requires either $\mathcal{O}(1)$, $\ll1$, or $\gg1$ coefficients.}%
\label{Fig2}%
\end{figure}

If the terms of the expansions~(\ref{HermQ}) are seen as nine-dimensional vectors, one can check that they span the space of complex (hermitian) matrices when the coefficients are arbitrary complex (real) numbers (in the squark case, four terms involving $(\mathbf{Y}_{i}^{\dagger}\mathbf{Y}_{i})^{2}$ have to be put back to get a basis)~\cite{ColangeloNS08}. Therefore, the first condition for MFV does not suffice to constrain the MSSM flavor structures; they can still be completely generic.

Let us imagine that the SUSY spectrum is fixed, e.g. from collider experiments. Flavor experiments then fix, or at least bound, the entries of the soft-breaking terms (so-called mass insertions) by constraining SUSY corrections like those in Eq.~(\ref{SUSYz}). If $M_{SUSY}$ is not much larger than $M_{W}$, these entries must be rather small, but there is no way to tell whether this is natural or not. Indeed, the naive definition of naturality --Lagrangian parameters should be $\mathcal{O}(1)$-- makes no sense in the flavor sector, where the known Yukawa couplings are already non-natural. To get a meaningful naturality principle, one should look at the coefficients of the expansions~(\ref{HermQ}), not the mass insertions. If we pass the experimental constraints on these coefficients, three different situations can arise (Fig.~\ref{Fig2})~\cite{MercolliS09}:\vspace{-0.1in}

\begin{itemize}
\item \textit{MFV flavor structure}: The second condition for MFV is for all coefficients to be natural, $a_{i},b_{i}\sim O(1)$. In that case, the soft-breaking terms inherit the hierarchies of the spurions. Consider for example $\mathbf{m}_{Q}^{2}$ in~(\ref{HermQ}). With $a_{i}\sim O(1)$, the leading non-diagonal effects are%
\begin{equation}
(\mathbf{m}_{Q}^{2})^{IJ}\approx m_{0}^{2}\;a_{2}\,(\mathbf{Y}_{u}^{\dagger}\mathbf{Y}_{u})^{IJ}\approx m_{0}^{2}\;a_{2}\,\,(m_{t}^{2}/v_{u}^{2})\,\,V_{3I}^{\ast}V_{3J}. \label{MFVpred}%
\end{equation}
Comparing Eq.~(\ref{SMz}) and (\ref{SUSYz}) with $m_{0}\sim M_{SUSY}$, the SUSY correction is tuned by the same CKM factor. If this suppression is necessary and sufficient for all FCNC processes, MFV solves the flavor puzzles.\vspace{-0.1in}

\item \textit{Fine-tuned flavor structure}: Some coefficients are required to be very small, $a_{i},b_{i}\ll1$. The suppression brought in by MFV is not sufficient, i.e. MFV fails to solve at least one flavor puzzle. In this case, one needs either $M_{SUSY}$ much larger than $1$ TeV, or a complementary/alternative fine-tuning mechanism for the flavor couplings.\vspace{-0.1in}

\item \textit{Generic flavor structure}: Some coefficients are required to be very large, $a_{i},b_{i}\gg1$, for instance if some FCNC processes deviate too much from their SM values. This could signal the presence of a new flavor structure. Indeed, though the terms of the expansions~(\ref{HermQ}) form a basis, they barely do so; they nearly live in a lower-dimensional subspace. Therefore, a flavor structure not aligned with the spurions can still be described by the expansions~(\ref{HermQ}), but requires huge coefficients.\vspace{-0.1in}
\end{itemize}

MFV thus offers an unambiguous test of naturalness. It permits to characterize flavor puzzles and to identify non-standard flavor structures. None of this is possible with the mass-insertion formalism.

\section{MFV phenomenology}

There are many studies of the phenomenological impacts of MFV. Here, only those based on the $G_{F}$ symmetry (see e.g.~\cite{CMFV} for another definition, or~\cite{Zwicky} for discrete group implementations) and in the context of the MSSM (see e.g.~\cite{DambrosioGIS02,ModIndep,CERN} for model-independent analyses) are briefly described.

\subsubsection*{Soft-breaking terms and CP-violation}

Before analyzing specific observables, there is one global issue to address. In its first implementations, MFV was assumed to commute with the CP-symmetry, so that only the CP-phases of the SM spurions would be present in the MSSM. However, there are three reasons why this cannot be true and MFV should allow for new CP-phases~\cite{MercolliS09}. First, CP-violation is a flavored phenomenon in the SM simply because there is no room for other phases (except for the QCD $\theta$-term). In the MSSM, there can be new CP-phases in flavor-blind sectors, beyond the reach of MFV. Second, the $G_{F}$ symmetry does not constrain CP-phases. All it asks is for soft-breaking terms to be expressed as in Eq.~(\ref{HermQ}) with free complex numbers as coefficients. Actually, the hermiticity of $\mathbf{m}_{Q,L}^{2}$ forces the coefficients in Eq.~(\ref{HermQ}) to be real, but this does not occur for trilinear terms, and further, the coefficients $a_{5}$, $b_{7-9}$, though real, are CP-violating. Finally, the very statement that $G_{F}$ and CP commute is incorrect. Indeed, $G_{F}$-singlet traces like $\langle(\mathbf{Y}_{u}^{\dagger}\mathbf{Y}_{u})^{n}(\mathbf{Y}_{d}^{\dagger}\mathbf{Y}_{d})^{m}...\rangle$ are understood to be included in the coefficients. Since they can be complex when the spurions are, CP cannot act separately on the expansion coefficients and on the spurions.

Generically, MFV is found compatible with current experimental constraints in the quark and lepton sectors. Assuming MFV is valid, visible deviations with respect to the SM are still possible in some $B$ but not in $K$ physics observables~\cite{CERN,Pheno}. However, the latter are ideal tests for MFV because being tuned by the small $V_{ts}^{\ast}V_{td}$ makes them particularly sensitive to new, non-minimal flavor structures. The additional CP-phases allowed by MFV are mostly felt in flavor-blind observables like electric dipole moments (EDM), but are still largely unconstrained once the SUSY and neutrino parameters are such that LFV transitions are not over-induced~\cite{MercolliS09,ParadisiS09}. Finally, it should be mentioned that if confirmed, the large CP-phase observed in $b\rightarrow s$ transitions~\cite{Bona} could not be accommodated within MFV~\cite{AltmannEtAl} and would indicate the presence of a flavor structure not aligned with $\mathbf{Y}_{u,d}$.

\subsubsection*{RGE behavior of MFV}

The MFV expansions are clearly invariant under the RGE, but to have the invariance of MFV requires in addition the naturality of the MFV coefficients at all scales. This is not automatic but depends on where MFV is first imposed. In the squark sector, starting with MFV at the GUT scale (enforced under $G_{F}$, see~\cite{MFVSU5} for an alternative), it survives at the low scale but in a very restrictive, and thus predictive, form. Indeed, ratios of coefficients show a fixed-point behavior~\cite{ColangeloNS08,RGE}, with all the CP-phases allowed by MFV at the GUT scale running towards zero at the low scale. This is mostly due to the fast QCD-driven running of flavor blind coefficients like $a_{1}$ in Eq.~(\ref{HermQ}), hence is not expected to hold in the slepton sector. 

The converse is not true. if MFV is imposed at the low scale, it holds at the GUT scale if and only if all the coefficients are close to their low-scale fixed points. Otherwise, some of them blow up at the GUT scale. In this case, MFV would need to derive from a relatively low-scale mechanism.

\subsubsection*{R-parity violating terms and proton stability}

How to enforce MFV on the RPV terms~(\ref{RPV}) follows from the same principles as for soft-breaking terms but for one difference in the treatment of the $U(1)$ factors of $G_{F}$, of which $U(1)_{\mathcal{B},\mathcal{L}}$ are linear combinations. To break them, MFV has to be enforced under $G_{F}^{\prime}\equiv SU(3)^{5}$. Note that $U(1)_{Q,L}$ should never be imposed since they are anomalous already in the SM~\cite{tHooft}. In practice, the only way to parametrize $\boldsymbol{\mu}^{\prime}$, $\boldsymbol{\lambda}  $, $\boldsymbol{\lambda}^{\prime}$, and $\boldsymbol{\lambda}^{\prime\prime}$ in a $G_{F}^{\prime}$-invariant way is to contract chains made of spurions by the $\varepsilon$-tensors of the $SU(3)$s, so as to explicitly break the flavor $U(1)$s and thereby either $U(1)_{\mathcal{B}}$ or $U(1)_{\mathcal{L}}$.

Typical proton decay bounds are extremely tight, e.g. $|\boldsymbol{\lambda}_{112}^{\prime}\boldsymbol{\lambda}_{112}^{\prime\prime}|\lesssim10^{-26}$ for squark masses around $300$ GeV~\cite{Barbier04}. To reach such a level, two mechanisms are at play in MFV~\cite{NikolidakisS07,ProcRPV}. First, the $\varepsilon$-tensor antisymmetry brings in light fermion mass factors, since one of its indices stands for the first generation. Second, the RPV couplings have peculiar properties under $G_{F}^{\prime}$. While $\mathbf{Y}_{u,d}$ are sufficient to parametrize the $\Delta\mathcal{B}$ coupling, all those violating $\mathcal{L}$ need the spurion $\boldsymbol{\Upsilon}_{\nu}$ of Eq.~(\ref{numass}), transforming as a \textbf{6} under $SU(3)_{L}$, and are thus proportional to the tiny $\nu_{L}$ masses.

These two mechanisms are enough to pass all proton decay bounds~\cite{NikolidakisS07}. The main motivation for R-parity thus disappears (it was already breached by higher dimensional operators~\cite{IbanezR91}). If we enforce MFV instead of R-parity, the phenomenological implications are profound. Besides proton decay, which could be close to current bounds, MFV predicts a sizeable $\Delta\mathcal{B}=1$, $t$-$s$-$d$ coupling (with anyone of the three being a squark), which could be observed at colliders through resonant stop production, enhanced top production, or LSP decays~\cite{NikolidakisS07}.

\section{Conclusion}

MFV, as a phenomenological hypothesis on the elementary flavor structures, is able to explain at once both the smallness of SUSY effects in FCNC and the very long proton lifetime. Both are direct consequences of the hierarchies of $\mathbf{Y}_{u,d,e}$ and of the tiny neutrino masses. MFV also offers a window into physics beyond the MSSM. First, it permits to identify those flavor couplings which are fine-tuned out of those which are just as ``natural'' as $\mathbf{Y}_{u,d,e}$. Within the MSSM, none of the flavor couplings currently needs to be fine-tuned, not even those violating R-parity. This ad-hoc symmetry is thus made redundant and can be avoided altogether. Second, CP-violation is not controlled by MFV. If the CP-phases present in the MSSM flavored or unflavored sectors are inducing too large effects, e.g. for EDM, they must be dealt with by some other mechanism. Third, the fact that a consistent picture emerges with only a few spurions points towards a rather simple origin for all the flavor structures. For all these reasons, the MFV hypothesis and its formalism are ideal to study the MSSM flavor structures. In addition, by allowing for a rich flavor sector compatible with experiment, MFV could soon prove central to SUSY searches at the LHC.

\end{document}